\documentclass[aps,preprint,amssymb,12pt,floatfix]{revtex4}
\usepackage{subfigure}
\usepackage{graphicx}
\usepackage{rotating}
\usepackage{tabularx}
\usepackage{epstopdf}

\def\rv{{\mathbf{r}}}
\def\Rv{{\mathbf{R}}}

\def\ds{{\Delta s}}
\def\la{\langle}
\def\ra{\rangle}
\def\xv{{\mathbf{x}}}
\def\Xv{{\mathbf{X}}}

\def\k{\kappa}
\def\b{\beta}

\begin{document}

\title{How accurate are polymer models in the analysis of FRET experiments on proteins?}

\renewcommand{\thefootnote}{\fnsymbol{footnote}}
\author{Edward P. O'Brien$^{1,2}$, Greg Morrison$^{1,3}$\footnote{E.O. and G.M. contributed equally to this work.}, 
Bernard R. Brooks$^{2}$, and D. Thirumalai$^{1,4}$\footnote{Corresponding author:
Institute for Physical Science and Technology, University of Maryland,
College Park, MD 20742, phone: 301-405-4803; fax: 301-314-9404;
e-mail: thirum@umd.edu}}
\affiliation{\small $^{1}$Biophysics Program\\
University of Maryland, College Park, MD 20742\\
$^{2}$Laboratory of Computational Biology\\
National Heart Lung and Blood Institute\\
National Institutes of Health, Bethesda, MD 20892\\
$^{3}$Department of Physics\\
University of Maryland, College Park, MD 20742\\
$^4$ Institute for Physical
Science and Technology\\
and Department of Chemistry and Biochemistry\\
University of Maryland, College Park, MD 20742\\
}
\vspace{2cm}

\begin{abstract} 
Single molecule Forster resonance energy transfer (FRET) experiments are used to infer the properties 
of the denatured state ensemble (DSE) of proteins.  From the measured average FRET efficiency, $\la E\ra$, the distance distribution $P(R)$ is inferred by assuming that the DSE can described as a polymer.  
The single parameter in the appropriate polymer model (Gaussian chain, Worm-Like chain, or Self-Avoiding walk)
for $P(R)$ is determined by equating the calculated and measured $\la E\ra$.  
In order to assess the accuracy of this ``standard procedure",
we consider the Generalized Rouse Model (GRM), 
whose properties ($\la E\ra$ and $P(R)$) can be analytically computed, and the Molecular Transfer 
Model for protein L for which accurate simulations can be carried out as a function 
of guanadinium hydrochloride (GdmCl) concentration. Using the precisely computed $\la E\ra$ 
for the GRM and protein L, we infer $P(R)$ using the standard procedure.
We find that the mean end-to-end distance can be 
accurately inferred (less than 10\% relative error) using $\la E\ra$ and polymer models for $P(R)$.  However,  
the value extracted for the radius of gyration ($R_g$) and the persistence length ($l_p$) are less accurate. 
For protein L, the errors in the inferred properties
increase as the GdmCl concentration increases for all polymer models.  
The relative error in the inferred $R_g$ and $l_p$, with respect to the exact values,
can be as large as 25\% at the highest GdmCl concentration. 
We propose a self-consistency test, 
requiring measurements of $\la E\ra$ by attaching dyes to different residues in the protein, to 
assess the validity of describing DSE using the Gaussian model.  Application of the
self-consistency test to the GRM shows that even for this simple model, which exhibits an order $\rightarrow$ 
disorder transition, the Gaussian $P(R)$ is inadequate. Analysis of experimental data of FRET efficiencies 
with dyes at several locations for the Cold Shock 
protein, and simulations results for protein L,  for which accurate FRET efficiencies 
between various locations were computed, shows that at high GdmCl concentrations 
there are significant deviations in the DSE $P(R)$ from the Gaussian model.  
\end{abstract} 

\date{\small \today} 
\maketitle
\baselineskip = 20pt

\newpage


\textbf{Introduction:} 
Much of our understanding of how proteins fold comes from experiments in which folding is initiated
from an ensemble of initially unfolded molecules whose structures are hard to characterize \cite{JacksonFD1998}. In many experiments, the initial
structures of the denatured state ensemble (DSE) are prepared by adding an excess amount of
denaturants or by raising the temperature above the melting temperature ($T_m$) of the protein \cite{FershtBook}.
Theoretical studies have shown that folding mechanisms
depend on the initial conditions, i.e. the nature of the DSE \cite{KlimovJMB2005}. Thus, a quantitative description
of protein folding mechanisms requires a molecular characterization of the DSE - a task that is made difficult
by the structural diversity of the ensemble of unfolded states \cite{EatonCOSB2008,HaranJPhys2003}.

In an attempt to probe the role of initial conditions on folding, single molecule FRET experiments 
are being used to infer the properties of unfolded proteins.
 The major advantage of these experiments 
is that they
can measure the FRET efficiencies of the DSE under solution conditions where the native state is stable.  The average denaturant-dependent FRET efficiency $\la E \ra$ has been used
to infer the global properties of the polypeptide chain in the DSE
as the external conditions are altered. The properties of the DSE are inferred from
$\la E \ra$ by assuming a polymer
model for the DSE, from which the root mean squared 
distance between two dyes attached at residues $i$ and $j$ along
the protein sequence ($R_{ij}={\la |\rv_i-\rv_j|\ra}$), the distribution of the end-to-end 
distance $P(R)$ (where $R={|\rv_N-\rv_0|}$), the root mean squared end-to-end distance 
($R_{ee}=\la \mathbf{R}^2\ra^{1/2}$),  the root mean squared radius of gyration ($R_g={\la \Rv_g^2\ra}^{\frac{1}{2}}$), and the 
persistence length ($l_p$) of the denatured protein \cite{WeissPNAS2000,HaasBIOCHEM2001,Haran2004,
UdgaonkarJMB2005,Nienhaus2005PNAS,Haran2006,Saxena2006JMB,SchulerPNAS2007,eatonPNAS2007,ThirumalaiCOSB1999} can be calculated. 

In FRET experiments, donor (D) and acceptor (A) dyes are attached at two locations
along the protein sequence \cite{NienhausMB2006,EatonCOSB2008}, and hence can only provide information about correlations between them. The efficiency of 
energy transfer $E$ between the D and A is equal to $(1+r^6/R_0^6)^{-1}$, where $r$
is the distance between the dyes, and $R_0$ is the dye-dependent  F\"{o}rster
distance \cite{NienhausMB2006,EatonCOSB2008}. Because of conformational fluctuations, there is a distribution
of $r$, $P(r)$, which depends on external conditions such as the temperature and denaturant concentration.
As a result, the average FRET efficiency $\la E \ra$ is given by
\begin{eqnarray}
\la E\ra &=&  \int_0^{\infty}(1+r^6/R_0^6)^{-1}P(r)dr, \label{fret}
\end{eqnarray}
under most experimental conditions, due to the central limit theorem \cite{SzaboJPCB2003}.  
If the dyes are attached to the ends of the chain, then $P(r)=P(R)$. 
Even if $\la E \ra$ is known accurately, the extraction of $P(R)$ from the integral equation (Eq. \ref{fret}) is 
fraught with numerical instabilities.  
In experimental applications to biopolymers, a functional form for $P(r)$ is assumed in order to satisfy the equality in Eq. \ref{fret}.  The form of $P(r)$ is based off of a particular polymer model which depends only on a single parameter (see Table \ref{models}):  The Gaussian chain (dependent on the Kuhn length $a$), the Wormlike Chain (WLC; dependent on the persistence length $l_p$), and the Self Avoiding Walk (SAW; dependent on the average end-to-end distance $R_{ee}$).  For the chosen polymer model meant to represent the biopolymer of interest, the free parameter ($a$, $l_p$, or $R_{ee}$) is determined numerically to satisfy Eq. \ref{fret}.  Using this method (referred to as the ``standard procedure" in this article), several researchers 
have estimated $R_g$ and $l_p$ as a function of the external conditions for protein L \cite{Haran2006,eatonPNAS2007}, Cold
Shock Protein (CspTm) \cite{SchulerPNAS2007}, and Rnase H \cite{NienhausMB2006}. The 
justification for using homopolymer models to analyze FRET data comes from the anecdotal comparison of the $R_g$ measured using
X-ray scattering experiments and the extracted $R_g$ from analysis of Eq. \ref{fret} \cite{EatonCOSB2008}.

Here, we study an analytically solvable generalized Rouse model (GRM) \cite{GRMpaper}
and the Molecular Transfer Model (MTM) for protein L \cite{ObrienPNAS2008} to assess the accuracy of using
polymer models to solve Eq. \ref{fret}. 
In the GRM, two monomers that are not covalently linked interact
through a harmonic potential that is truncated at a distance $c$. The presence of the additional length scale, $c$, which
reflects the interaction between non-bonded beads, results
in the formation of an ordered state as the temperature $(T)$ is varied.  A more detailed discussion of these models can be found in the Methods section.  
For the GRM, $P(R)$ can be
analytically calculated, and hence the reliability of the standard
procedure to solve Eq. \ref{fret} can be unambiguously established.  
We find that the accuracy of the polymer models in extracting the exact
values in the GRM depends on the location of the monomers that are constrained by the
harmonic interaction.  Using coarse-grained simulations of protein L, we show that the error
between the exact quantity and that inferred using the standard procedure depends on the property of interest.
For example,  the 
inferred end-to-end distribution $P(R)$ is in qualitative, but not quantitative agreement with the exact $P(R)$ distribution
obtained from accurate simulations. 
In general, the DSE of protein L is better characterized by the SAW polymer model than the Gaussian chain model.

We propose that the accuracy of the popular Gaussian model can be
assessed by measuring $\la E \ra$ with dyes attached at multiple sites
in a protein \cite{SchmidJMB2006,UdgaonkarJMB2007,SchulerPNAS2007}.
If the DSE can be described by a Gaussian chain, then the parameters extracted by attaching
the dyes at position $i$ and $j$ can be used to predict $\la E \ra$ for dyes at other
points. The proposed  self-consistency test shows that the
Gaussian model only qualitatively accounts for the experimental data 
of CspTm, simulation results for protein L, and the exact analysis of the GRM.

\begin{figure}[tbp]
\begin{center}
\includegraphics[width=.9\textwidth]{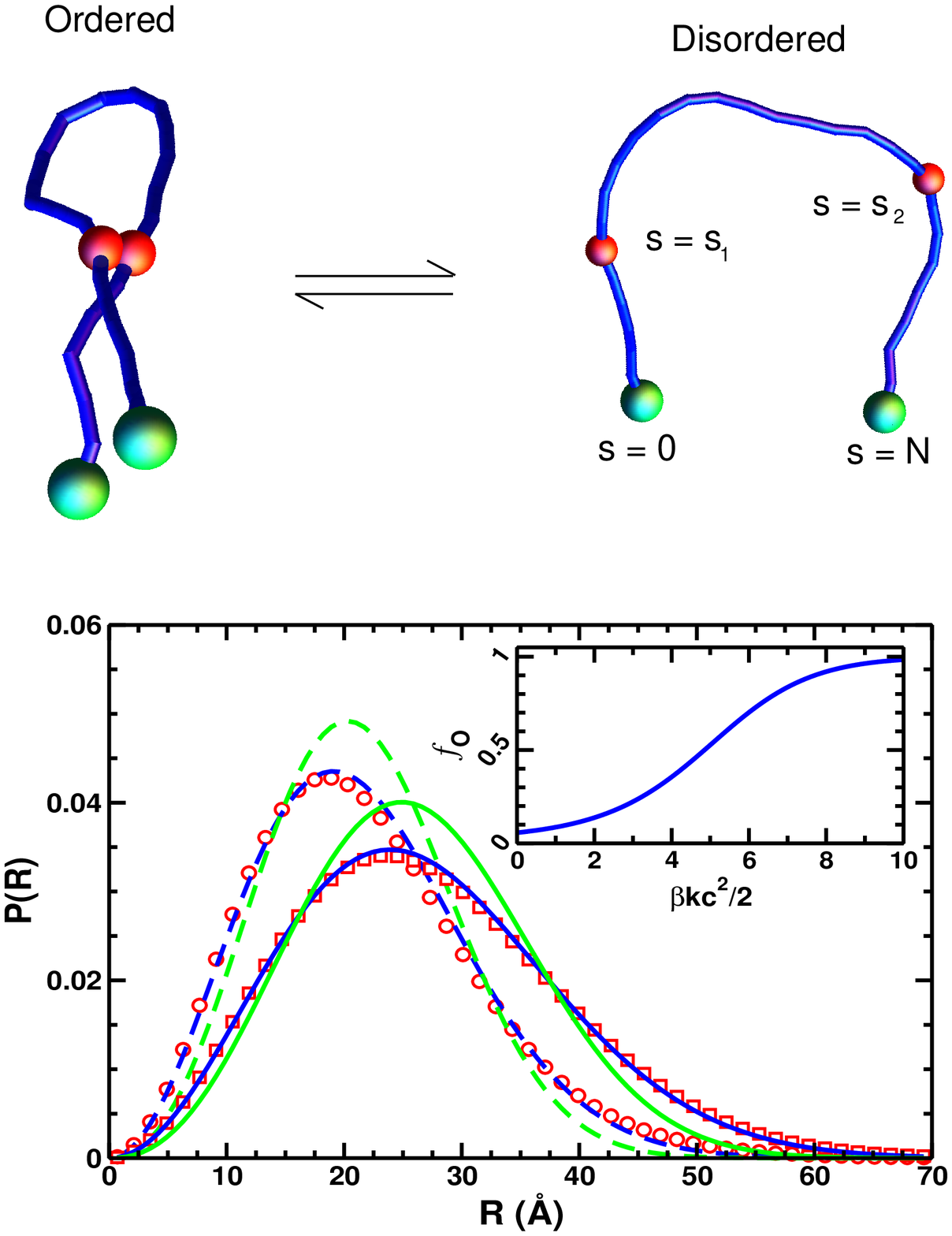}
\caption{}
\label{Fig1.fig}
\end{center}
\end{figure}

\noindent
\textbf{\large{Results and Discussion}}
\\
\\
We present the results in three sections. In the first and second sections we examine the
accuracy of the standard procedure (described in the introduction) in accurately inferring the properties of the denatured state
of the GRM and protein L models. The third section presents results of the Gaussian
Self-consistency Test applied to these models. We also analyze experimental data
for CspTm to assess the extent to which the DSE deviates from a Gaussian chain.
\\
\\
\noindent
\textbf{I. GRM:} 
The Generalized Rouse model (GRM) is a simple modification of the Gaussian chain with $N$ bonds and Kuhn length $a_0$, which includes a single, non-covalent bond between two monomers at positions $s_1$ and $s_2$ (Fig. \ref{Fig1.fig}).  The monomers at $s_1$ and $s_2$ interact with a truncated harmonic potential with spring constant $k$, with strength $\k= kc^2/2$, where $c$ is the distance at which the interaction vanishes (Eq. \ref{GRMPot}).  The GRM minimally represents a two state system, with a clear demarcation between ordered (with $|\rv(s_2)-\rv(s_1)|\le c$) and disordered (with $|\rv(s_2)-\rv(s_1)|> c$) states. Unlike other polymer models 
(see Table I), which are characterized by a single length scale, the GRM is 
described by  $a_0$ and the energy scale $\k$.  For $\beta \k\to0$ (the 
high temperature limit, where $\beta=1/k_B T$), the simple Gaussian chain is 
recovered (see Methods for details).  By varying $\b\k$, a disorder $\to$ order 
transition can be induced (see Fig. \ref{Fig1.fig}).  The presence of the interaction
between monomers $s_1$ and $s_2$ approximately mimics persistence of structure 
in the DSE of proteins.  If the fraction of ordered states, $f_O$, exceeds 0.5 
(Fig. \ref{Fig1.fig} inset), we assume that the residual structure is present 
with high probability.  The exact analysis of the GRM when $|\rv(s_2)-\rv(s_1)|\le c$ 
allows us to examine the effect of structure in the DSE on the global properties 
of unfolded states.  

Because 
$\la E\ra$ can be calculated exactly for the GRM (see Eq. \ref{grmAv}), it can be used to quantitatively study the accuracy of solving Eq. \ref{fret} using the standard procedure
\cite{WeissPNAS2000,Nienhaus2005PNAS,Haran2006,SchulerPNAS2007,eatonPNAS2007}.
Given the best fit for the Gaussian chain (Kuhn length $a$), WLC (persistence length $l_p$), and SAW (average end-to-end distance $R_{ee}$), as described in Table \ref{models}, many quantities of interest can be inferred ($P(R)$ or $R_g$, for example), and compared with the exact results for the GRM.
The extent to which the exact and inferred properties deviate, due to the additional single energy scale in the GRM, is an indication of the accuracy of the standard procedure used to analyze Eq. \ref{fret}.
\\
\\
{\bf{$P(R)$ is accurately inferred using the Gaussian polymer model}}:  
If the interacting monomers are located near the endpoints of the chain, 
the end-to-end distribution function is bimodal, with a clear distinction 
between the ordered and disordered regions \cite{GRMpaper}.  However, 
if the monomers $s_1$ and $s_2$ are in the interior of the chain, 
the two-state behavior is obscured because the distribution function becomes unimodal.
In Fig. \ref{Fig1.fig}, we show the exact and inferred $P(R)$ functions for a chain with 
$N=63$, $a_0=3.8$\AA, $c=2a_0$, and $|s_2-s_1|=(N-1)/2=31$.   We take the F\"orster distance 
(Eq. \ref{fret}) $R_0=23$\AA$ \lesssim{\la{\mathbf{R}}^2\ra^{1/2}_{\k=0}}$ for the GRM.  
The distributions are unimodal for both weakly ($\b\k=2$) and strongly ($\b\k=6.6$) interacting monomers.  

The strength of the interaction is most clearly captured with the fraction of conformations in the ordered state, $f_O$, with $f_O =0.25$ 
for the weakly interacting chain and $f_O = 0.75$ for the strongly interacting chain (inset of Fig. \ref{Fig1.fig}).  
The inferred Gaussian distribution functions are in excellent agreement with the exact result.   
Because of the underlying Gaussian Hamiltonian in the GRM, 
the rather poor agreement in the inferred SAW distribution seen in Fig \ref{Fig1.fig} 
is to be expected.  We also note that the GRM is inherently flexible, so that the 
WLC and Gaussian chains produce virtually identical distributions.  
\\
\\
{\bf{The accuracy of the inferred $R_g$ depends on the location of the interaction}}:  
The two-state nature of the GRM is obscured by the relatively long unstructured regions of the chain, similar to the effect seen in laser optical tweezer experiments with flexible handles \cite{GRMpaper}.  As a result, $P(R)$ is well represented by a Gaussian chain, with a smaller inferred Kuhn length, $a\le a_0$ (Fig. \ref{Fig2.fig}).   
For large $\b \k$, where the ordered state is predominantly occupied and $\rv(s_2)\approx\rv(s_1)$, the end-to-end distribution function is well approximated by a Gaussian chain with $N^*=N-\Delta s$ bonds.  Consequently, the single length scale for the Gaussian chain, decreases to $a\sim a_0\sqrt{1-\Delta s/N}\approx 0.71a_0$ for large values of $\b \k$ (Fig. \ref{Fig2.fig}).   

Because the two-state nature of the chain is obscured for certain values of $|s_2 - s_1|$, 
the Gaussian chain gives an excellent approximation to the end-to-end 
distribution function.  However, the radius of gyration $R_g$ is not as accurately 
obtained using the Gaussian chain model, as shown in Fig. \ref{Fig3.fig}.  The exact $R_g$ 
for the GRM reflects both the length scale $a_0$ and the energy scale $\b \k$, which can 
not be fully described by the single inferred length scale $a$ in the Gaussian chain.  
For the GRM, $R_g$ depends not only on the separation between the monomers $\Delta s$, 
but also explicitly on $s_1$ (i.e. where the interaction is along the chain; see 
Fig. \ref{Fig3.fig} and the Methods section), which can not be captured by the 
Gaussian chain.   If the interacting monomers are in the middle of the chain 
($s_1=(N+1)/4=16$ and $\Delta s=31$), the inferred $R_g$ is in excellent 
agreement with the exact result (Fig. \ref{Fig3.fig}).  The relative error 
in $R_g$ (the difference between the inferred and exact values, divided 
by the exact value) 
is no less than -2\%.  
However, for interactions near the endpoint of the chain, with $s_1=0$ and the same $\Delta s = 31$, 
the relative error between the inferred and exact values of $R_g$ is $\sim -14$\%.  
The large errors arise because  the radius of gyration depends on the behavior 
of all of the monomers, so that the energy scale $\b \k$ plays a much larger role in the determination of $R_g$ than  $R_{ee}$.
\\
\\
\noindent
\textbf{II. MTM for protein L:}
Protein L is a 64 residue protein (Fig. \ref{pl_efret_pairs}A) whose folding has been studied by a variety of methods \cite{BakerFD1997,Baker1999Nature,bakerJMB2000,Haran2006,eatonPNAS2007}. More recently, single molecule FRET experiments have been used to probe changes in the
DSE as the concentration of GdmCl is increased from 0 to 7 M \cite{Haran2006,eatonPNAS2007}.
From the measured GdmCl-dependent $\la E \ra$, the properties of the DSE, such 
as $R_{ee}$, $P(R)$, and $R_g$, were extracted by solving Eq. \ref{fret}, and assuming a Gaussian
chain $P(R)$ \cite{Haran2006,eatonPNAS2007}. 
To further
determine the accuracy of polymer models in the analysis of $\la E\ra$, we use
simulations of protein L in the same range of the concentration of denaturant, [C], as used in experiments \cite{WeissPNAS2000,UdgaonkarJMB2005}.
\\
\\
\noindent
\textbf{The average end-to-end distance is accurately inferred from FRET data:}
In a previous study \cite{ObrienPNAS2008}, we showed that the predictions
based on MTM simulations for protein L are in excellent agreement with 
experiments. 
From the calculated 
$\la E\ra$ with the dyes at the endpoints (solid black line in Fig. \ref{pl_efret_pairs}B), which is in quantitative agreement with
experimental measurements \cite{ObrienPNAS2008}, we determine the model parameter
$R_{ee}$ or $l_p$ by assuming that the exact $P(R)$ can be
approximated by the three polymer models in Table \ref{models}.  Comparison of the exact value of $R_{ee}$ to the inferred value $R_F$, obtained using the
simulation results for $\la E \ra$, shows
good agreement for all three polymer models (Fig. \ref{error}A). There are deviations
between $R_{ee}$ and $R_F$ at [C] $>$ C$_m$, the midpoint of the 
folding transition. The maximum relative error (see inset of Fig. \ref{error}A)
we observe is about 10\% at the highest concentration of GdmCl.
The SAW model provides the most accurate estimate of $R_{ee}$ at GdmCl
concentrations above C$_m$, with a relative error $\leq 0.05$, and the
Gaussian model gives the least accurate values, with a relative error $ \leq 0.10$ (Fig. \ref{error}A). Due to
the relevance of excluded volume interaction in the DSE of real proteins, the better
agreement using the SAW is to be expected.
\\
\\
\noindent
\textbf{Polymer models do not give quantitative agreement with
the exact $P(R)$:} 
The inferred distribution
functions, $P_F(R)$'s, obtained by the standard procedure (as described in the introduction) at [C]=2 M and 6 M GdmCl
differ from the exact results (Fig. \ref{error}B).  Surprisingly, the agreement between $P(R)$ and
$P_F(R)$ is worse at higher [C]. The range of $R$ explored and the width of the  exact
distribution are less than predicted
by the polymer models. The Gaussian chain and the SAW models account only for chain entropy,
while the WLC only models the bending energy of the protein. However,
in protein L (and in other proteins) intra-molecular attractions
are still present even when [C]=6 M $>$ C$_m$. As a result,
the range of $R$ explored in the protein L simulations 
is expected to be less than in these polymer models.
Only at [C]/C$_m >> 1$ and/or at high $T$ are proteins expected
to be described by Flory random coils. Our results show that although it is possible to use
models that can give a single quantity correctly ($R_{ee}$, for example),
the distribution functions are less accurate. The results in Fig. \ref{error}B
show that $P(R)$, inferred from the polymer models, agrees only qualitatively with the exact
$P(R)$, with the SAW model being the most accurate (Fig. \ref{error}B).  While the MTM will not perfectly reproduce all of the fine details of Protein L under all situations, we expect it to produce more realistic results than idealized polymer models, which have no specific intra-chain interactions.
\\
\\
\noindent
\textbf{Inferred $R_g$ and $l_p$ differ significantly from the exact values:}
The solution of Eq. \ref{fret} using a Gaussian chain or WLC model yields
$a$ and $l_p$, from which $R_g$ can be analytically calculated (Table \ref{models}).
Figs. \ref{error2}A and \ref{error2}B, which compare the FRET inferred $R_g$ and $l_p$
with the corresponding values obtained using MTM simulations, show that the relative errors
are substantial. At high [C] values the $R_g^F$ deviates from $R_g$ by nearly 25\% if the 
Gaussian chain model is used (Fig. \ref{error2}A). The value of $R_g \approx$ 26 \AA\ at [C]$=8$ M
while $R_g^F$ using the Gaussian chain model is $\approx 31$ \AA. 
In order to obtain reliable estimates of $R_g$, an accurate
calculation of the distance distribution between all the heavy atoms in a protein is needed.
Therefore, it is reasonable to expect that errors in the inferred $P(R)$
are propagated, leading to a poor estimate of internal distances, thus resulting in a larger error in $R_g$.   
A similar inference can be drawn about the persistence length obtained using polymer models (Fig. \ref{error2}B).
Plotting $l_p^{F}$ as a function of [C] (Fig. \ref{error2}B), against $l_p=R_{ee}/2L$, shows that $l_p$ is overestimated 
at concentrations above 1 M GdmCl, with the error increasing as [C] increases. 
The error is less when the Gaussian chain model is used.
\\
\\
\noindent
\textbf{III. Gaussian Self-consistency test shows the DSE is non-Gaussian:}
The extent to which the Gaussian chain accurately describes the ensemble of
conformations that are sampled at different values of the external
conditions (temperature or denaturants) can be assessed by performing
a self-consistency test.
A property of a Gaussian chain is that if the average root mean square distance, $R_{ij}$, between two monomers
$i$ and $j$ is known then $R_{kl}$, the distance between
any other pair monomers $k$ and $l$, can be computed using
\begin{eqnarray}
R_{kl} &=& \sqrt{\frac{|k-l|}{|i-j|}}R_{ij}. \label{gscc_eq}
\end{eqnarray}
Thus, if the conformations of a protein (or a polymer) can be modeled
as a Gaussian chain, then $R_{ij}$ inferred from the FRET
efficiency $\la E_{ij}\ra$ should
accurately predict $R_{kl}$ and the FRET efficiency
$\la E_{kl} \ra$, if the dyes were to be placed at monomers $k$ and $l$.  We refer to this criterion as the Gaussian self-consistency (GSC) test,
and the extent to which the predicted $R_{kl}$ from Eq. \ref{gscc_eq} deviates from the
exact $R_{kl}$ reflects
deviations from the Gaussian model description of the DSE. 

{\it GRM:} 
For the GRM, with a non-bonded interaction
between monomers $s_1$ and $s_2$, we calculate $\la E_{ij}\ra$ using Eq. \ref{grm_fret} with
$j$ fixed at 0 and for $i= 20, 40,$ and 60. Using the exact results for
$\la E_{ij}\ra$, the values of $R_{ij}$ are inferred assuming that $P(r)$
is a Gaussian chain. 
From the inferred $R_{ij}$ the values of $\la E_{kl} \ra$ and $R_{kl}$ can be calculated using Eqs.  \ref{fret} and \ref{gscc_eq}, respectively.  
We note that, since $R_{kl}/R_{ij}=\sqrt{|k-l|/|i-j|}$ (Eq. \ref{gscc_eq}) for any pair $(k,l)$ using the Gaussian chain model, the prediction of the Gaussian chain will be independent of the particular choices of $k$ and $l$, as long as their difference is held constant.  We first apply the GSC test to a GRM in which $f_O\approx 0.75$ due to a
favorable interaction between monomers $s_1=16$ and $s_2=47$.
There are discrepancies between the values of the Gaussian inferred ($R_{kl}^G$) 
and exact $R_{kl}$ distances, as well as the inferred ($\la E_{kl}^G\ra$) and 
exact $\la E_{ij} \ra$ efficiencies when a Gaussian model is used (Fig. \ref{GrmGsccFig}).
The relative errors in the predicted values of the
FRET efficiency and the inter-dye distances can be as large as 30-40\%,
depending on the choice of $i$ and $j$ (see insets in Fig. \ref{GrmGsccFig}).  We note that the relative error in the end-to-end distance is small for dyes near the endpoints (the green line in Fig. \ref{GrmGsccFig}b), in agreement with the results shown in Fig. \ref{Fig1.fig}.  The errors decrease as $f_O$ decreases,
with a maximum error of 20\% when $f_O=0.5$, and 10\% when $f_O=0.25$
(data not shown). By construction, the GRM is a Gaussian chain when $f_O=0$ and
therefore the relative errors will vanish at sufficiently small $\beta \k$ (data not shown).
These results show that even for the GRM, with only one non-bonded interaction 
in an otherwise Gaussian chain, its DSE cannot
be accurately described using a Gaussian chain model.
Thus, even if the overall end-to-end distribution
$P(r)$ for the GRM is well approximated as a Gaussian (as seen in Fig. \ref{Fig1.fig}),
the internal $R_{kl}$ monomer pair distances can deviate from predictions
of the Gaussian chain model.

{\it Protein L:}  We apply the GSC test to our simulations of protein L at GdmCl concentrations of [C]=2.0 M (below C$_m$=2.4M)
and [C]=7.5 M (well above C$_m$). While our simulations allow us to compute the DSE $\la E_{ij}\ra$ for
all possible $(i,j)$ pairs, we examine only a subset of $\la E_{ij}\ra$ 
as a function of GdmCl concentration (Fig. \ref{pl_efret_pairs}B).  By choosing multiple $j$ values for the same value of $i$, we can determine whether distant residues along the backbone are close together spatially, which may offer insights into three-point correlations in denatured states.  We note that all values of $\la E_{ij}\ra$ in Fig. 4 are monotonically decreasing, except for the (1,14) pair.  This is due to the fact that the native state has a beta-strand between these two residues; as the protein denatures, they come closer together, increasing the FRET efficiency.  We use these values for $\la E_{ij}\ra$ in the GSC test. The results are shown in Figs. \ref{gscc_pl}A and \ref{gscc_pl}B.  Relative errors in $\langle E_{kl}\rangle$ as large as 36\% at 2.0 M GdmCl and 50\% at 7.5 M GdmCl are found, with the lowest errors generally seen for residues close to one another along the backbone, in agreement with the results from the GRM (Fig. 7a inset).  In addition, the number of data points that underestimate $\la E_{kl}\ra$ increases
as [C] is changed from 7.5 M to 2.0 M for $|k-l| < 20$. Despite these differences,
the gross features in Figs. \ref{gscc_pl}A and \ref{gscc_pl}B are concentration independent.
Because the error does not vanish for all $(k,l)$ pairs (Figs. \ref{gscc_pl}A and \ref{gscc_pl}B),
we conclude that the DSE of protein L cannot be modeled as a Gaussian chain.
\\
\\
\noindent
\textbf{The GSC test for CspTm:}
In an interesting single molecule experiment, Schuler and coworkers have
measured FRET efficiencies by attaching donor and acceptor dyes to pairs
of residues at five different locations of a CspTm \cite{SchulerPNAS2007}. They analyzed the data
by assuming that the DSE properties can be mimicked using a Gaussian
chain model. We used the GSC test to predict $\la E_{kl}\ra$ for
dyes separated by $|k-l|$ along the sequence using
the experimentally measured values $\la E_{ij}\ra$.

The relative error in $\la E_{kl}\ra$ (Eq. \ref{gscc_eq}) should be zero 
if CspTm can be accurately modeled as a Gaussian chain.
However, there are significant deviations (up to 17\%) 
between the predicted and experimental values (Fig. \ref{gscc}).
The relative error is fairly insensitive to 
the denaturant concentration (compare Figs. \ref{gscc}A and \ref{gscc}B).  
It is interesting to note that the trends in Fig. \ref{gscc} are qualitatively similar to the relative errors in the GRM 
at $f_O > 0$.
Based on these observations we conclude tentatively that whenever the DSE is ordered
to some extent (i.e., when there is persistent residual structure) then we expect deviations
from a homopolymer description of the DSE of proteins. At the very least, the GSC
test should be routinely used to assess errors in the modeling of the DSE as a Gaussian chain.
\\
\\
\textbf{\large{Conclusions}}
\\
\\
In order to assess the accuracy of polymer models to infer the properties of the DSE of 
proteins from measurement of FRET efficiencies, we studied two models for which 
accurate calculations of all the equilibrium  properties can be carried out.  Introduction 
of a non-bonded interaction between two monomers in a Gaussian chain (the GRM) leads to 
an disorder-order transition as the temperature is lowered.  The presence of `residual structure' 
in the GRM allows us to 
clarify its role in the use of the Gaussian chain model to fit the accurately calculated FRET 
efficiency.  Similarly, we have used the MTM model for protein L to calculate precisely the 
denaturant-dependent $\la E\ra$ from which we extracted the global properties of the DSE 
by solving Eq. \ref{fret} using the $P(R)$'s for the polymer models in Table I.  Quantitative comparison of the 
exact values of a number of properties of the DSE 
(obtained analytically for the GRM and accurately using simulations for protein L)
and the values inferred from 
$\la E\ra$ has allowed us to assess the accuracy with which polymer 
models can be used to analyze the experimental data.  
The major findings and implications of our study are listed below.

(1)  The polymer models, in conjunction with the measured $\la E\ra$, can accurately infer 
values of $R_{ee}$, the average end-to-end distance.  However, $P(R)$, $l_p$, and $R_g$ are not 
quantitatively reproduced.  For the GRM, $R_g$ is underestimated, whereas it is 
overestimated for protein L.   The simulations show that the absolute value of the relative error in the inferred 
$R_g$ can be nearly 25\% at elevated GdmCl concentration.

(2)  We propose a simple self consistency test to determine the ability of the Gaussian chain model to 
correctly infer the properties of the DSE of a polymer.  Because the Gaussian chain depends only on a 
single length scale, the FRET efficiency can be predicted for varying dye positions once $\la E\ra$ 
is accurately known for one set of dye positions.  The GSC test shows that neither the GRM, simulations
of  protein L, nor experimental data on CspTm can be accurately modeled using the Gaussian chain.  
The  relative errors between the exact and 
predicted FRET efficiencies can be as high as 50\%.  For the GRM, we find that the variation in 
the FRET efficiency as a function of the dye position changes abruptly if one dye is placed near 
an interacting monomer.   
Taken together these findings suggest that it 
is possible to infer the structured regions in the DSE by systematically varying the location of the dyes.  This is due to the fact that the FRET efficiency is perfectly monotonic using the Gaussian Chain model.  An experiment that shows non-monotonic behavior in $\la E_{ij}\ra$ as the dye positions $i$ and $j$ are varied is a clear signal of non-Gaussian behavior, and sharp changes in the FRET efficiency as a function of $|i-j|$ may indicate strongly interacting sites (see Fig. \ref{GrmGsccFig}a).

(3)  The properties of the DSE inferred from Eq. \ref{fret} become increasingly more 
accurate as [C] decreases.  At a first glance this finding may be  surprising, especially considering that stabilizing
intra-peptide interactions are expected to be weakened at high GdmCl concentrations 
[C], and therefore the protein should be more ``polymer-like."  The range of $R$-values 
sampled at low [C] is much smaller than at high [C].  Protein L swells as [C] is 
increased, as a consequence of the increase in the solvent quality.  It is possible that 
[C]$\approx$2.4 M might be close to a $\Theta$-solvent (favorable intrapeptide and solvent-peptide 
interactions are almost neutralized), so that $P(R)$ can be 
approximated by a polymer model.  The inaccuracy of polymer models in describing 
$P(R)$ at [C]$=$6 M suggests that only at much higher concentrations does protein L behave as 
a random coil.  In other words, $T$=327.8 K and [C]$=$6 M is not an athermal (good) solvent.  

(4)  It is somewhat surprising that polymer models, which do not have side chains or any preferred 
interactions between the beads, are qualitatively correct in characterizing the DSE of 
proteins with complex intramolecular interactions.  In addition, even [C]=6 M GdmCl is not 
an athermal solvent, suggesting that at lower [C] values the aqueous denaturant may be 
closer to a $\Theta$-solvent.  A consequence of this observation is that, for many 
globular proteins, the extent of collapse may not be significant, resulting in the 
nearness of the concentrations at which collapse and folding transitions occur, as shown
by Camacho and Thirumalai \cite{Camacho1993PNAS} some time ago.  We suggest that only 
by exploring the changes in the conformations of polypeptide chains over a wide range 
of temperature and denaturant concentrations can one link the variations of the 
DSE properties (compaction) and folding (acquisition of a specific structure).
\\
\\
\noindent
\textbf{\large{Theory and computational methods}}
\\
\\
\textbf{GRM model:}
In order to understand the effect of a single non-covalent interaction between two monomers along a chain, we consider a Gaussian chain with Kuhn length $a_0$ and $N$ bonds, with a harmonic attraction between monomers $s_1 \le s_2$, which is cutoff at a distance $c$.  The Hamiltonian for the GRM is
\begin{eqnarray}
\beta H&=&\frac{3}{2a^2}\int_0^N ds \ \dot\rv^2(s)+\beta V[\rv(s_2)-\rv(s_1)]\label{GRMHam}\\
\beta V[\rv]&=&\left\{\begin{array}{cc}k\rv^2/2 & |\rv|<c \\ kc^2/2 & |\rv|\ge c \end{array}\right. \label{GRMPot},
\end{eqnarray}
where $k$ is the spring constant that constrains $\rv(s_2)-\rv(s_1)$ to a harmonic well.  The Hamiltonian in Eq. \ref{GRMHam} allows the exact determination of many quantities of interest.  Defining $\xv=\rv(s_2)-\rv(s_1)$ and $\ds=s_2-s_1$, we can determine most averages of interest for the GRM using
\begin{eqnarray}
\la\cdots\ra&=&\frac{\int d^3\rv_1 d^3\xv d^3\rv_N(\cdots)G(\xv,\rv_N;\ds,N)}{\int d^3\rv_1 d^3\xv d^3\rv_N\ G(\xv,\rv_N;\ds,N)}\label{grmAv}\\
G(\xv,\rv_N;\ds,N)&=&\exp\bigg(-\frac{3\xv^2}{2\ds\, a^2}-\frac{3(\rv_N-\xv)^2}{2(N-\ds)a^2}-\beta V[\xv]\bigg).
\end{eqnarray}

\textbf{$C_\alpha$-SCM protein model and GdmCl denaturation:} 
We use the coarse-grained
$C_{\alpha}$-side chain model ($C_{\alpha}$-SCM) to model protein L 
(for details see the supporting information in \cite{ObrienPNAS2008}).
In the $C_{\alpha}$-SCM each residue in the polypeptide chain is
represented using two interaction sites, one that is centered on the
$\alpha$-carbon atom and another that is located at the center-of-mass of the
side chain \cite{thirumPNAS2000}. Langevin dynamics simulations \cite{thirumalaiFD1997} are carried out
in the underdamped limit at zero molar guanidinium chloride. Simulation details
are given in \cite{ObrienPNAS2008}.

We model the denaturation of protein L by GdmCl
using the molecular transfer model (MTM) \cite{ObrienPNAS2008}. MTM combines simulations
at zero molar GdmCl with experimentally measured
transfer free energies, using a reweighting method \cite{Swendsen1989PRL,KumarJCC1992,SheaJCP1998} to predict the equilibrium
properties of proteins at any GdmCl concentration of interest.  
\\
\\
\noindent
\textbf{Analysis:}
\\
\\
{\emph{GRM:}}
The average squared end-to-end distance can be computed directly from Eq.
\ref{grmAv}, using $\la\Rv_{ee}^2\ra=Na_0^2+(\la\xv^2\ra-\ds \,a_0^2)$.
The exact expression for $\la\xv^2\ra$ is easily determined, but somewhat
lengthy, and we omit the explicit result here.  Also of interest is the
end-to-end distribution function, $P(\Rv{})=\la\delta[\rv_N-\Rv{}]\ra$,
which can be obtained from Eq. \ref{grmAv}.  In order to determine the
probability of an interior bond being in the `ordered' state (i.e. the fraction
of residual structures, see the inset for Fig. \ref{Fig1.fig}a), we compute
the interior distribution, $P_I(\Xv)=\la \delta[\xv-\Xv]\ra$, so that
$f_{O}=\int_{|\xv|\le c}d^3\xv \, P_I(\xv)$.  The radius of gyration requires
a more complicated integral than the one found in Eq. \ref{grmAv}, but we find
\begin{eqnarray}
R_g^2 =\frac{N a_0^2}{6}+(\la\xv^2\ra-\ds\, a_0^2)\bigg[\frac{\ds}{3N}+\frac{s_1}{N}-\bigg(\frac{\ds}{2N}+\frac{s_1}{N}\bigg)^2\bigg]
\end{eqnarray}
Note that, unlike the average end-to-end distance, the radius of gyration depends not only on $\ds$, but also on $s_1$.

The FRET efficiency for a system with dyes attached to $\rv(j=0)=\mathbf{{0}}$ and $\rv(i)$, $\la E\ra=\la[1+(|\rv(i)|/R_0)^6]^{-1}\ra$, is determined from Eq. \ref{grmAv} as
\begin{eqnarray}
E(i)=\left\{\begin{array}{cc}E^{G}(i) & 0\le i\le s_1 \\\frac{\int_0^\infty dxdr\, g_1(x,r;\{s_i\})/[1+(r/R_0)^6]}{\int_0^\infty dxdr\, g_1(x,r;\{s_i\})} & s_1<i<s_2 \\ \frac{\int_0^\infty dxdr\, g_2(x,r;\{s_i\})/[1+(r/R_0)^6]}{\int_0^\infty dxdr\, g_2(x,r;\{s_i\})}& s_2\le i\le N\end{array}\right.  \label{grm_fret}
\end{eqnarray}
where $E{}^{G}(i)$ is the FRET efficiency for a Gaussian chain with $i$ bonds, and
\begin{eqnarray}
g_1(x,r;\{s_i\})&=&xr\ \sinh\bigg(\frac{3(i-s_1)x r}{\lambda a_0^2}\bigg)e^{-3(i x^2+\ds r^2)/2\lambda a_0^2-\beta V[x]}\\
g_2(x,r;\{s_i\})&=&xr\ \sinh\bigg(\frac{3 x r}{(i-\ds)a_0^2}\bigg) e^{-3x^2/2\ds a_0^2-3(x^2+r^2)/2(i-\ds)a_0^2-\beta V[x]}\\
\lambda&=&(s_2+s_1)i-s_1^2-i^2
\end{eqnarray}
This result allows us to compute the Gaussian Self-consistency test, after a numerical integral over $r$.
\\
\\
\emph{Protein L:} 
Averages and distributions were computed using the
MTM \cite{ObrienPNAS2008} which combines experimentally measured transfer free
energies \cite{BolenBIOCHEM2004}, converged simulations and the WHAM equations \cite{Swendsen1989PRL,KumarJCC1992,SheaJCP1998}.
The WHAM equations use the simulation time-series of potential energy
and the property of interest at various
temperatures and gives a best estimate of the averages and distributions
of that property. The native state ensemble (NSE) and DSE
subpopulations were defined as having a structural RMSD (root mean squared deviation),
after least squares minimization, of less than or greater than 5 \AA\ relative to the crystal structure for
the NSE and DSE respectively. 
The exact values of $l_p$ are computed using the average $R$ from simulations and the
relationships listed in Table \ref{models}.
\\
\\
{\bf{Notation:}}
Throughout the paper, exact values of all quantities are reported without superscript or subscript.  For the GRM, exact values are analytically obtained or calculated by performing a one-dimensional integral numerically.  For convenience, exact results for protein L refer to converged simulations.  While these simulations have residual errors, the simplicity of the MTM has allowed us to calculate all properties of interest with arbitrary accuracy.  The use of subscript or superscript is, unless otherwise stated, reserved for quantities that are extracted by solving Eq. \ref{fret} using the polymer models listed in Table I.
\\
\\
\noindent
\textbf{Acknowledgments:}
We thank Sam Cho, Govardan Reddy, and David
Pincus for their comments on the manuscript.
E.O. thanks Guy Ziv for many useful discussions on experimental aspects of 
FRET measurements and analysis. 
This work was supported in part by grants from the NSF (05-14056) to D.T., 
a NIH GPP Biophysics Fellowship to E.O., by the Intramural
Research Program of the NIH, National Heart Lung and Blood Institute.

\begin{sidewaystable}
\centering
\caption{Polymer models and their properties}
\vspace{1cm}
\renewcommand{\thefootnote}{\thempfootnote}
\begin{tabularx}{220mm}{lc|cc|cc|c}
\hline
 && \multicolumn{5}{c}{Property} \\
\hline
Polymer Model && End-to-end distribution $P(R)$\footnote{The average end-to-end distance $R_{ee}=\left(\int R^2P(R)dR\right)^{1/2}$} && Radius of gyration $R_g$ && Persistence length $l_p$ \\
\hline
Gaussian && $4\pi R^2\left(\frac{3}{2\pi Na^2}\right)^{3/2} \exp\left(\frac{-3R^2}{2Na^2}\right)$ && $a\sqrt{N/6}$ && $\frac{Na^2}{2L}=\frac{a}{2}$\\
Worm-like Chain\footnote{$L$ and $l_p$ are the contour length and persistence length respectively. 
$C_1 = (\pi^{3/2}e^{-\alpha}\alpha^{-3/2}(1+3\alpha^{-1}+\frac{15}{4}\alpha^{-2}))^{-1}$ where $\alpha = 3L/(4l_p)$. $C_2 = 1/(2l_p)$.} && 
$\frac{4\pi R^2C_1}{L(1-(R/L)^2)^{9/2}}\exp\left(\frac{-3L}{4l_p (1-(R/L)^2)}\right)$ && $\frac{L}{6C_2} 
+ \frac{1}{4C_2^2} + \frac{1}{4LC_2^3} - \frac{1-\exp(- L/l_p)}{8 C_2^4 L^2}$ && 
$R_{ee}^2 = 2l_pL-2l_p^2-2l_p^2\exp(-\frac{L}{l_p})$\footnote{Using the simulated $\la R^2\ra$, $l_p$ was solved
for numerically using this equation.}\\
Self Avoiding Polymer\footnote{$\theta$ and $\delta$ equal 0.3 and 2.5, respectively. The constants
$a$ and $b$ are determined by solving the integrals of the zeroth and second moment of $\int P(R)dr = \int R^2P(R) dr =1$, resulting in values of $a=3.67853$ and $b=1.23152$.} && 
$\frac{a}{R_{ee}}(\frac{R}{R_{ee}})^{2+\theta}\exp(-b\left(\frac{R}{R_{ee}}\right)^{\delta})$ && N/A && N/A \\
\hline
\end{tabularx}
\label{models}
\end{sidewaystable}

\newpage
\begin{center}
\textbf{\large{Figure Captions}}
\end{center}

\textbf{Figure 1}:
Top figures shows a schematic sketch of the GRM, with the donor and acceptor at the 
endpoints, represented by the green spheres, and the interacting monomers at $s_1$ 
and $s_2$ represented by the red spheres.  In the ordered configuration, the monomers 
at $s_1$ and $s_2$ are tightly bound. The bottom figure shows the exact and the inferred 
end-to-end distribution functions $P(r)$ for interior interactions ($\Delta s=31$).  
The blue lines correspond to the Gaussian chain model, light green lines to the SAW, 
and the symbols to the exact GRM distribution.  Dashed lines and red circles are 
for $\b \k=6.6$, while solid lines and red squares correspond to $\b \k=2$.
In the inset we show the fraction of ordered states as a function of $\b \k$.  Note 
that 75\% of the structures are ordered at $\b \k=6.6$, yet the inferred Gaussian 
$P(r)$ is in excellent agreement with the exact result.

\textbf{Figure 2}:
The inferred Kuhn length $a$ as a function of $\b\k$ for the GRM.  $R_{ee}$ monotonically 
decreases a function of the interaction strength, leading to the decrease in $a/a_0$.  
The Kuhn length $a$ reaches its limiting value of $a\approx a_0\sqrt{1-\Delta s/N}$ when $f_O\approx 1$.

\textbf{Figure 3}:
Comparison of the exact (symbols) and inferred (blue line) values of the radius of gyration ($R_g$) 
as a function of $\b \k$ for $\Delta s=31$.  Shown are $R_g$'s for the GRM with $s_1=0$ 
(open symbols) and $s_1=16$ (filled symbols) for $N=63$.  The structures in the ordered 
state are shown schematically.  The $R_g$ obtained using the standard procedure is independent of 
$s_1$, while the exact result is not.  The inset shows the relative errors between the inferred and exact values of $R_g$.

\textbf{Figure 4}:
(a) A secondary structure representation of protein L in its native state.
Starting from the N-terminus, the residues are numbered 1 through 64. (b) The average FRET
efficiency between the various $(i,j)$ residue pairs in protein L versus GdmCl concentration.
The $\la E_{ij} \ra$ values, computed using MTM simulations,  for each $(i,j)$ pair is 
indicated by the two numbers next to each line. For example, the
numbers `1-64' beneath the black line indicates that $i=1$ and 
$j=64$.  The solid black line (lowest values of $\la E\ra$) is computed for the dyes at the endpoints.

\textbf{Figure 5}:
(a) The root mean squared end-to-end distance ($R_{ee}$) as a function of  GdmCl concentration
for protein L. The average $R_{ee}$ (black circles) and the $R$ for the sub-population of the DSE (red squares) 
from simulations are shown.
The values of  $R_{ee}$ inferred by solving Eq. (1) by the standard procedure using the Gaussian chain, Worm Like Chain, 
and Self Avoiding polymer models
are shown for comparison as the top, middle and bottom solid lines respectively. 
The inset shows the relative error between the exact and the values inferred using the FRET efficiency for $R_{ee}$ versus
GdmCl concentration. The top, middle and bottom lines correspond to the Gaussian chain, Worm Like Chain and
Self Avoiding Walk polymer models respecitvely.
(b) Simulation results of the denatured state end-to-end distance distribution ($P(R)$)
at 2.4 M GdmCl (solid red squares) and 6 M GdmCl (open red squares) and $T$=327.8 K are compared with
$P(R)$s  using the Gaussian chain, Worm Like Chain, and Self Avoiding Walk
polymer models are also shown at 2.4 M GdmCl (dashed lines) and 6 M GdmCl (solid lines).
The top middle and bottom lines correspond to the Self Avoiding Walk, Worm Like Chain, and Gaussian chain
polymer models.

\textbf{Figure 6}:
(a) Comparison of $R_{g}$ from direct simulations of protein L
and that obtained by solving Eq. (1) using the Gaussian chain, 
and Worm Like Chain polymer models. The top line (magenta) shows the WLC fit, 
the bottom line (blue) shows the Gaussian fit, red squares show the DSE $R_g$ from 
the simulation, and black circles show the average simulated $R_g$. 
The inset shows the relative errors as a function of GdmCl concentration;
top and bottom lines correspond to the Gaussian chain and Worm Like Chain
polymer models respectively.
(b) Same as (a) except the figure is for $l_p$. Top and bottom lines 
correspond to the inferrred $l_p$ using the Gaussian chain and Worm Like Chain polymer models respectively.
Top and bottom sets of squares correspond to a direct analysis
of the simulations using the Worm Like Chain and Gaussian chain 
polymer models respectively.
  
\textbf{Figure 7}:
Gaussian Self-consistency test using (a) the FRET efficiency and (b) the average end-to-end distance
for the GRM with $f_O = 0.75$ and interaction sites at $s_1=16$ and $s_2=47$.  In both (a) and (b) the solid lines are the inferred
properties and the open symbols are the exact values.
In both (a) and (b), $j=0$ and the blue, magenta, and green lines correspond
to a dye at $i=20, 40$, and 60, respectively.
The insets show the relative error for $\la E_{kl}\ra$ and $R_{kl}$. Note that
the relative error would be zero if the Gaussian chain accurately
modeled the GRM.

\textbf{Figure 8}: The Gaussian self consistency test applied to
simulated DSE $\la E_{ij}\ra$ data of protein L using the $(i,j)$
pairs listed in Fig. \ref{pl_efret_pairs}B.
Shown are the relative errors at (a) 2.0 M GdmCl and (b) 7.5 M GdmCl.
In both (a) and (b), solid green circles
correspond to $|i-j|= 13$, open orange squares to $|i-j|=16$, blue
squares to $|i-j|=19$, open brown circles to $|i-j|=29$, cyan
$\ast$ to $|i-j|=30$, red diamonds to $|i-j|=34$, solid violet
triangles to $|i-j|=44$, open grey triangles to $|i-j|=50$, and
magenta x's to $|i-j|=54$.  The color of each point corresponds to the color
of each line in Fig. \ref{pl_efret_pairs}b, except for the 1-64 pair,
which is not shown here.
 
\textbf{Figure 9}:
The Gaussian Self-consistency test (GSC) using experimental data
from CspTm. One dye was placed at one endpoint, and the
location of the other was varied.  We show relative error of the predicted
$\la E\ra$, using Eqs. \ref{fret} and \ref{gscc_eq}, versus the distance
between the dyes ($|k-l|$) for [C]=2M (a) and 5M (b).  In both (a) and (b), triangles
correspond to $|i-j|= 33$, x's to $|i-j|=45$, diamonds to $|i-j|=46$, squares to $|i-j|=57$, and circles to $|i-j|=65$. The trends in Figs. (7) and(8) are similar.

\newpage

\begin{figure}[tbp]
\begin{center}
\includegraphics[width=.9\textwidth]{grm_kuhn_length_vs_temp.eps}
\caption{}
\label{Fig2.fig}
\end{center}
\end{figure}

\begin{figure}[tbp]
\begin{center}
\includegraphics[width=.8\textwidth]{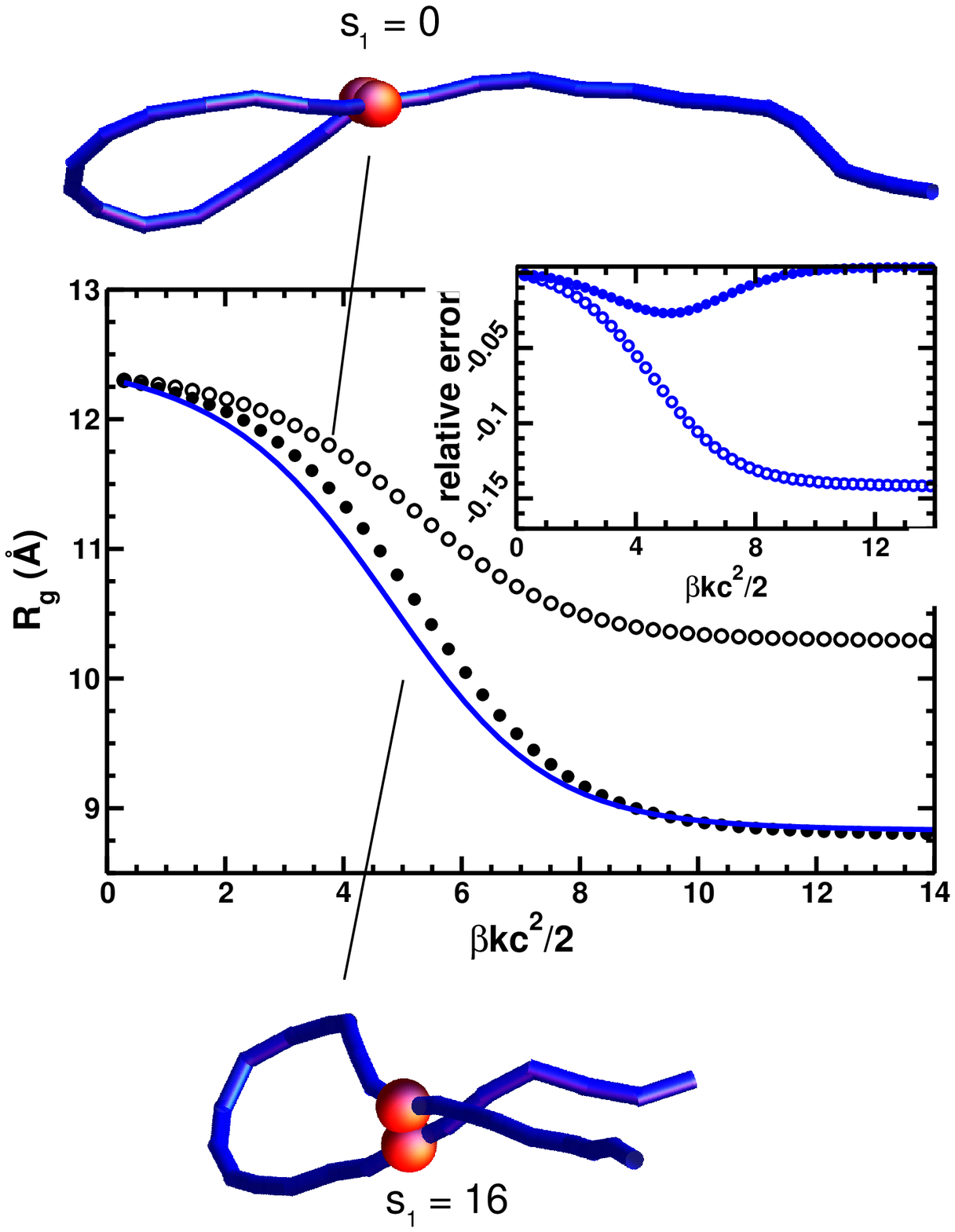}
\caption{}
\label{Fig3.fig}
\end{center}
\end{figure}

\begin{figure}[htb]
\begin{center}
\subfigure[]{
  \label{}
  \includegraphics[width=.25\textwidth]{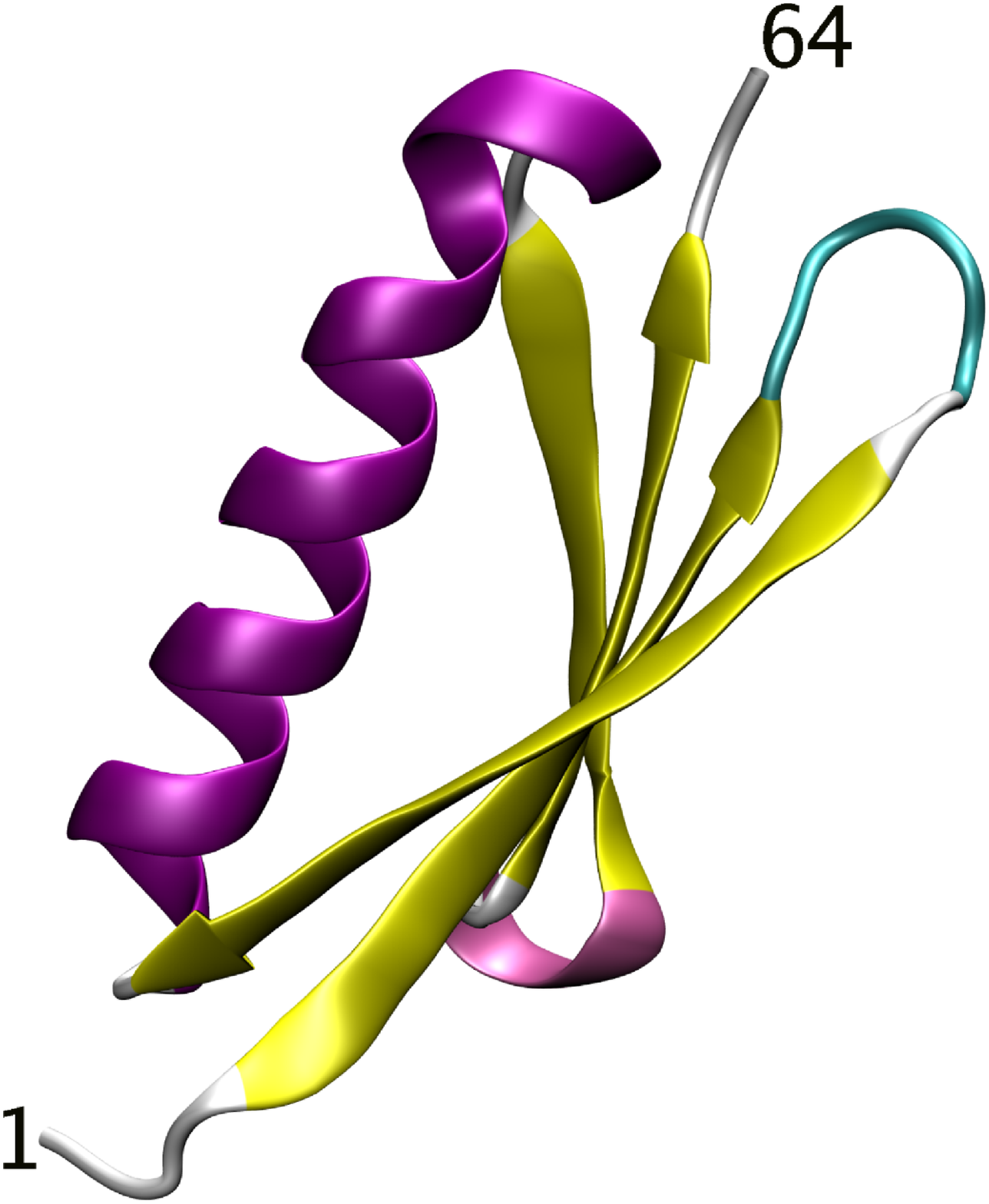}
  }\\
\vspace{1cm}
\subfigure[]{
  \label{}
  \includegraphics[width=0.7\textwidth]{efret_vs_gdcl_various_pairs_327.8k.eps}
  }
\caption{}
\label{pl_efret_pairs}
\end{center}
\end{figure}

\begin{figure}[ht]
\subfigure[]{
  \label{}
  \includegraphics[width=3.8in]{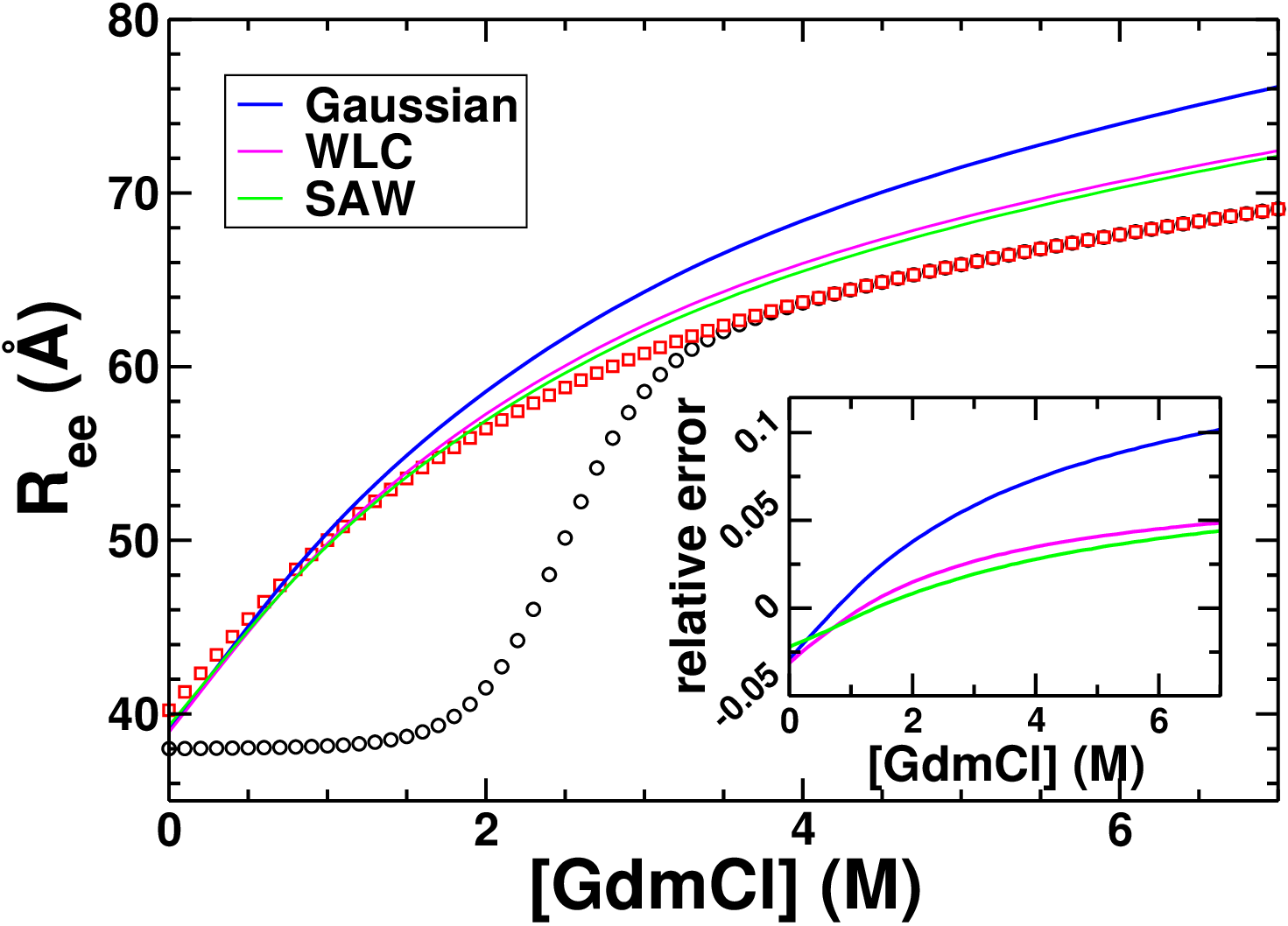}
  }\\
\vspace{1cm}
\subfigure[]{
  \label{}
  \includegraphics[width=3.8in]{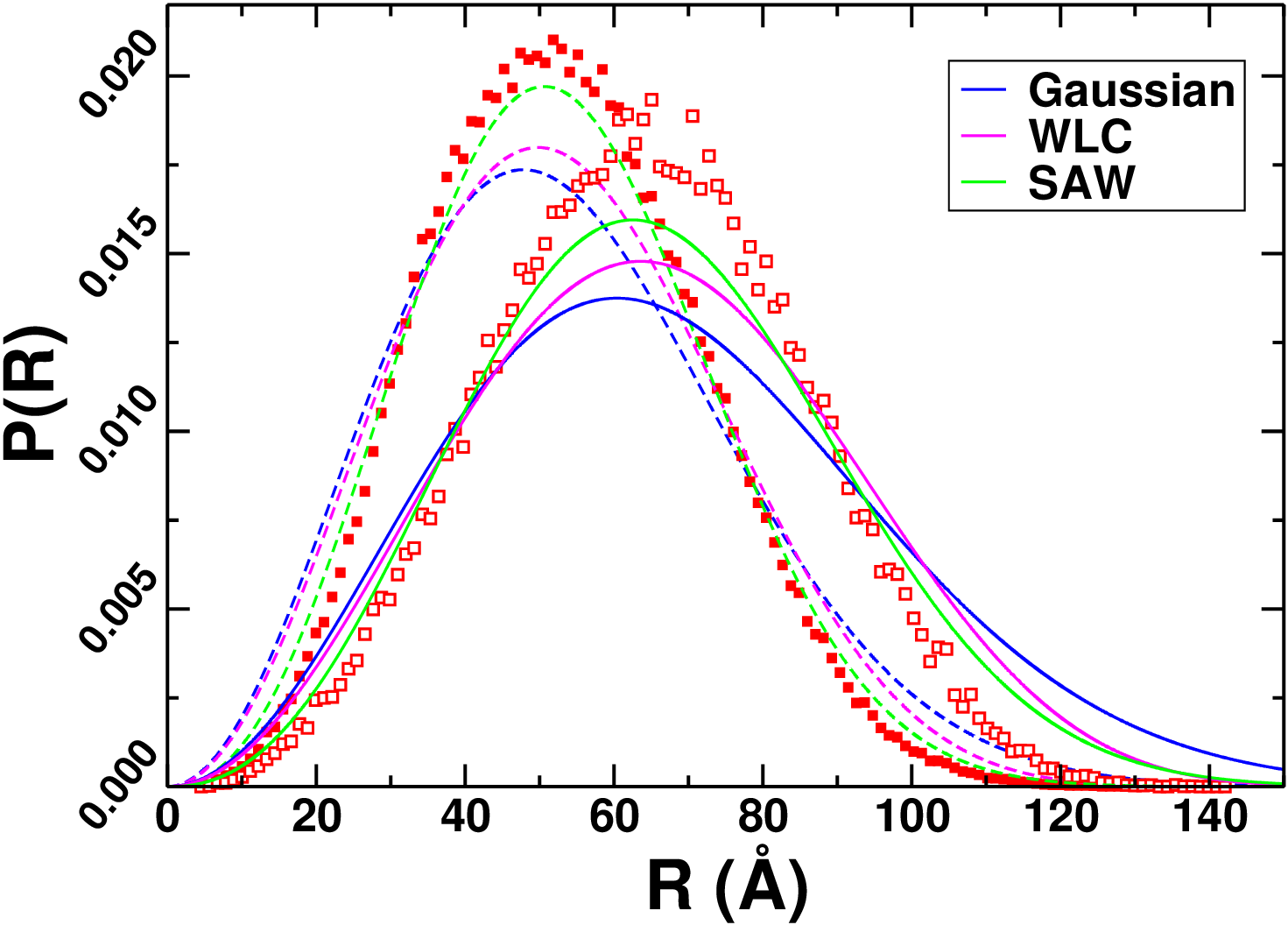}
  } 
\caption{}
\label{error}
\end{figure}

\begin{figure}[ht]
\subfigure[]{
  \label{}
  \includegraphics[width=3.8in]{rg-dse_vs_conc_327.8k_3.eps}
  }\\
\vspace{1cm}
\subfigure[]{
  \label{}
  \includegraphics[width=3.8in]{lp-dse_vs_conc_327.8k_3.eps}
  }
\caption{}
\label{error2}
\end{figure}

\begin{figure}[htbp]
\begin{center}
\includegraphics[width=.7\textwidth]{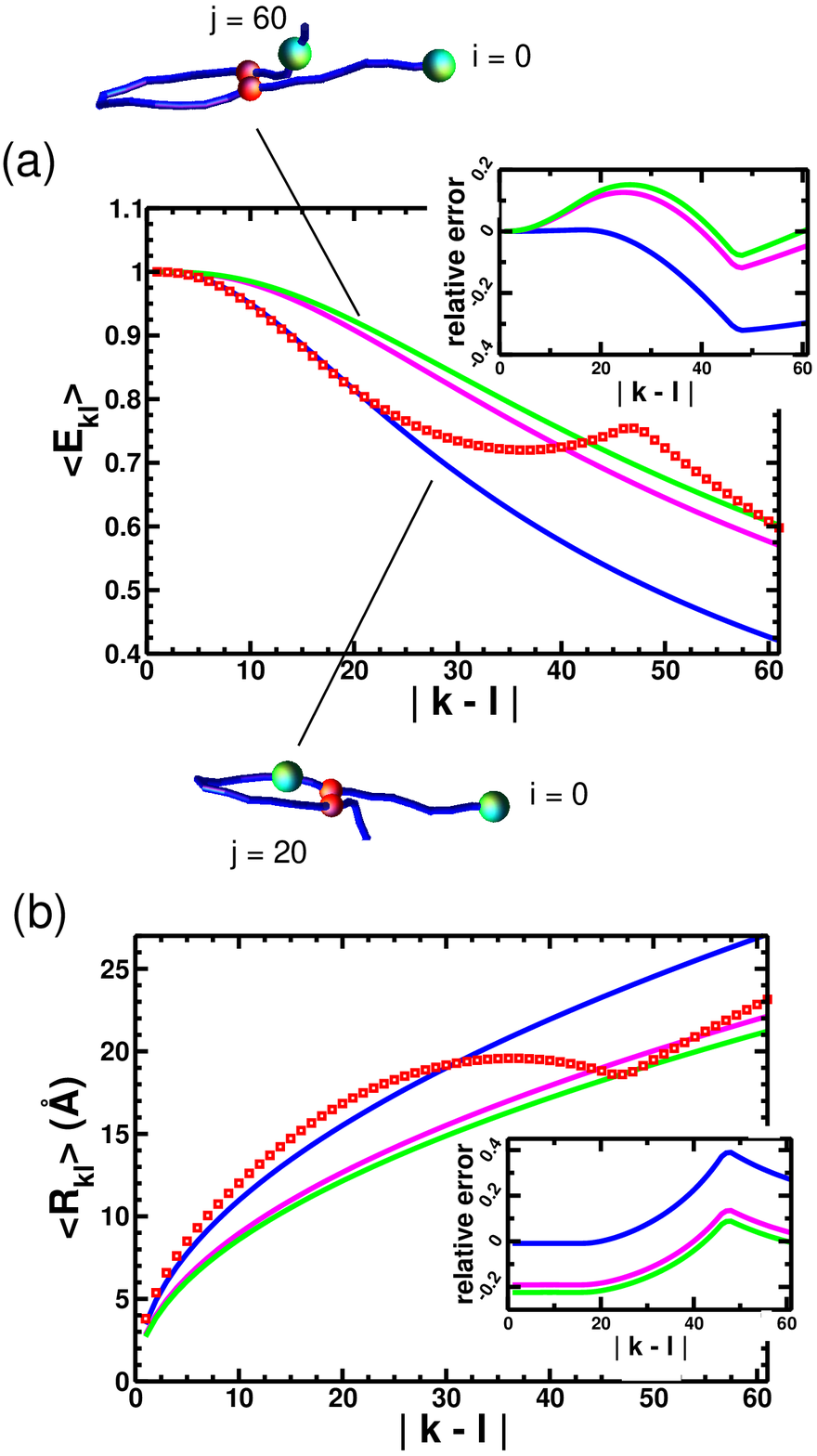}
\caption{}
\label{GrmGsccFig}
\end{center}
\end{figure}

\begin{figure}[htb]
\subfigure[]{
  \label{}
  \includegraphics[width=3.8in]{pl_gscc_ro55_2.0m_gdcl_327k_3.eps}
  }\\
\vspace{1cm}
\subfigure[]{
  \label{}
  \includegraphics[width=3.8in]{pl_gscc_ro55_7.5m_gdcl_327k_3.eps}
  }
\caption{}
\label{gscc_pl}
\end{figure}

\begin{figure}[ht]
\subfigure[]{
  \label{}
  \includegraphics[width=3.8in]{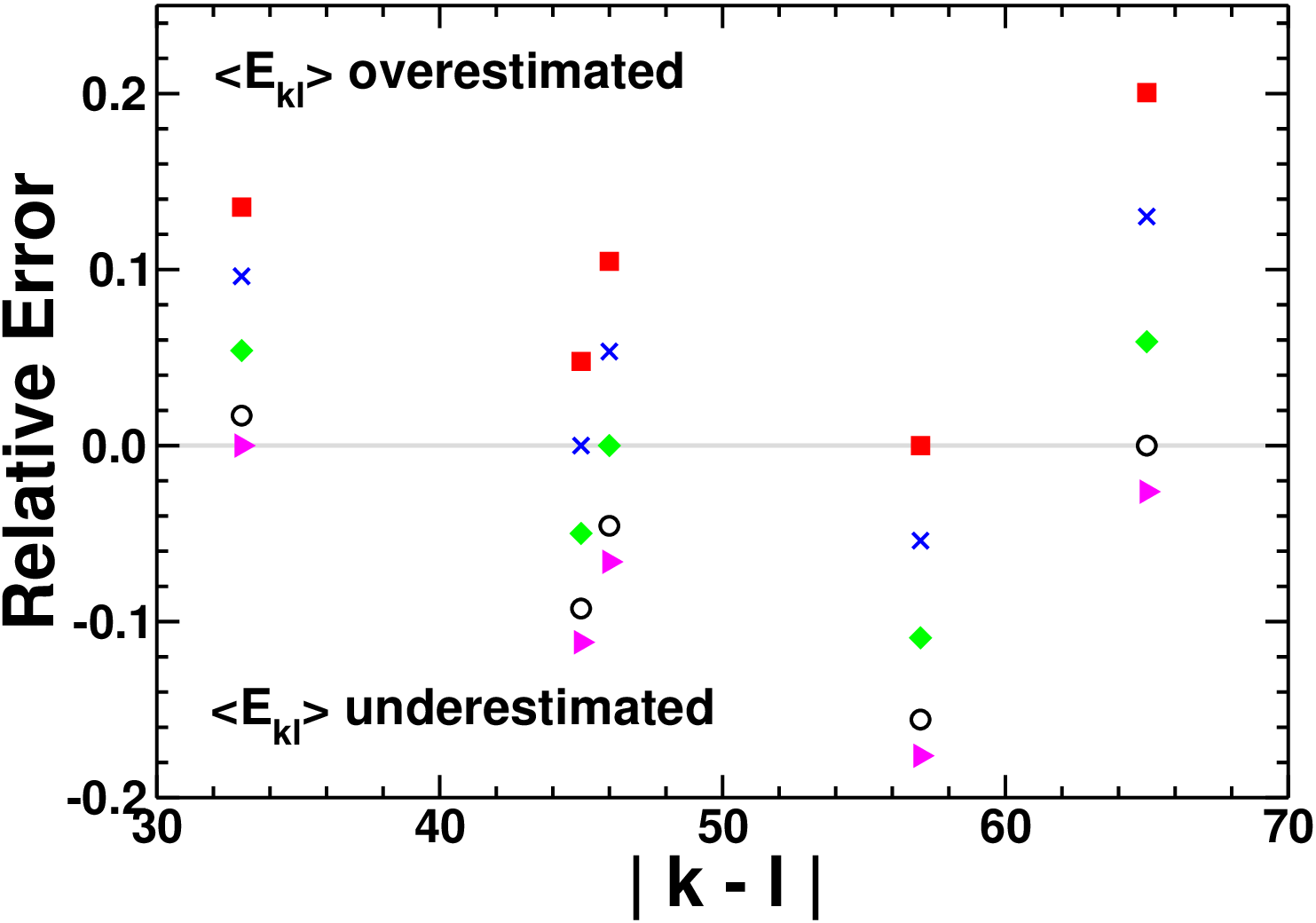}
  }\\
\vspace{1cm}
\subfigure[]{
  \label{}
  \includegraphics[width=3.8in]{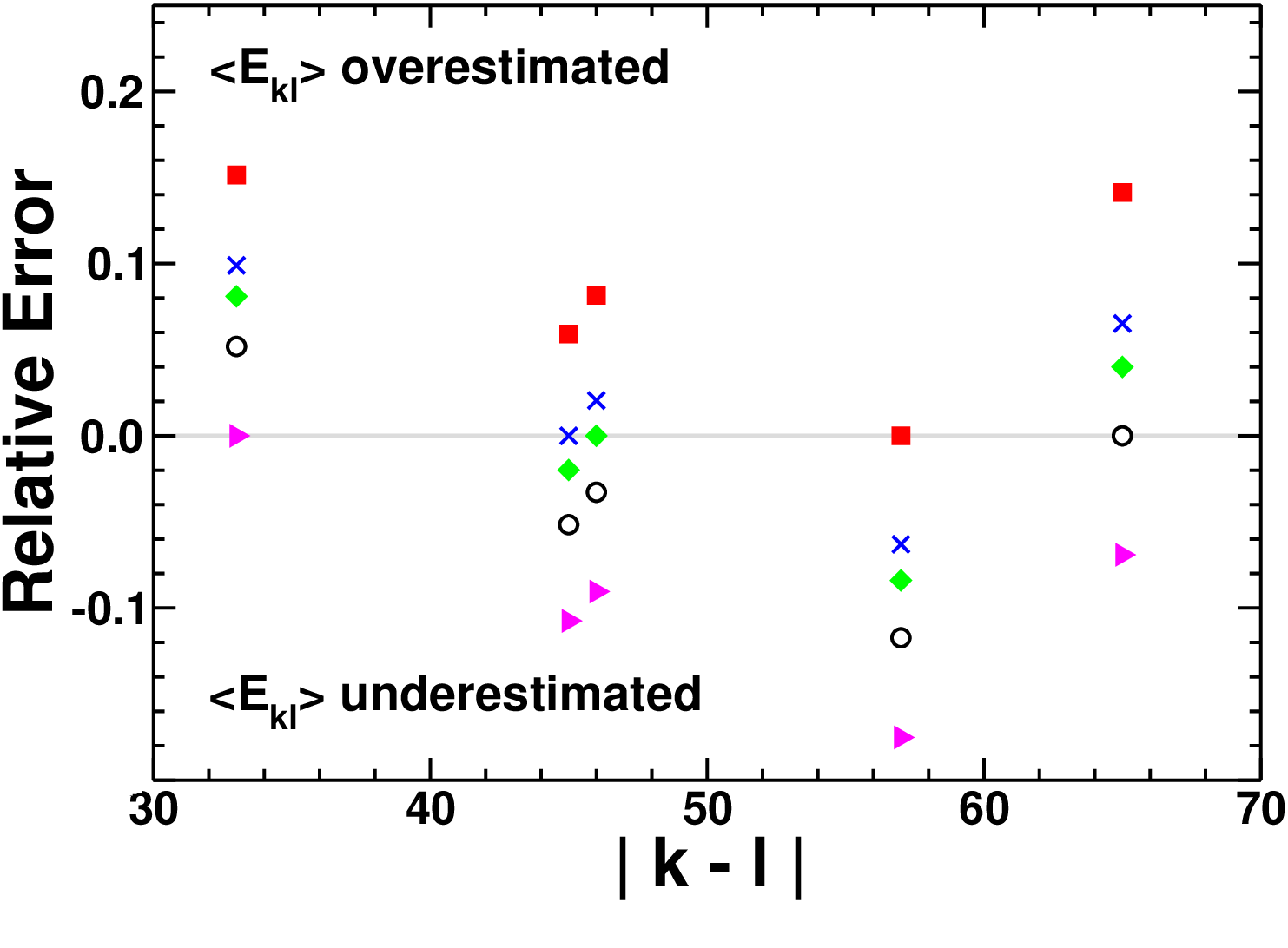}
  }
\caption{}
\label{gscc}
\end{figure}

\end{document}